# Hot Carrier Mobility Dynamics Unravel Competing Sub-ps Cooling Processes in Lead Halide Perovskites

Andrés Burgos-Caminal, Juan Manuel Moreno-Naranjo, Aurélien René Willauer,
Arun Abi Paraecattil, Ahmad Ajdarzadeh, and Jacques-E. Moser *

Photochemical Dynamics Group, Institute of Chemical Sciences and Engineering,
École Polytechnique Fédérale de Lausanne, 1015 Lausanne, Switzerland

* Corresponding author, email: je.moser@epfl.ch

## Abstract

The outstanding photovoltaic performances of lead halide perovskite (LHP) thin film solar cells are due in particular to the large diffusion length of photocarriers. The mechanism behind this property and its dependence on the various anions and cations in the LHP material composition are still an active topic of debate. Here, we apply ultrabroadband terahertz spectroscopy with a time resolution ≤ 200 fs to probe the early carrier mobility dynamics of photoexcited perovskite samples of different chemical compositions. An increase in the carrier mobility with time constants in the range of 0.3 to 0.7 ps is observed for different LHPs, in which the signal amplitude is larger at shorter excitation wavelengths. This feature is assigned to the relaxation of hot carriers from low-curvature regions of the band structure to the bottom of the valley, where their effective mass accounts for the maximum mobility. The dependence of the mobility dynamics on the initial photogenerated carrier population shows that carrier cooling competes with a dynamic screening effect associated with polaron formation and relaxation. A kinetic model that considers all competing processes is proposed. Simulations based on this model allow us to fit the experimental data well and evidence a composition dependence of the dynamic screening lifetime.

## Introduction

Lead halide perovskites (LHPs) have attracted much attention in recent years due to the outstanding photovoltaic performances of thin film solar cells based on these materials combined with their facile processability.[1,2] These advantages have encouraged many groups to seek a detailed understanding of the charge carrier and quasiparticle dynamics that govern the observed performances. Two of the main factors behind the large power conversion efficiencies are generally agreed to be a) slow recombination of photogenerated charge carriers[3] and b) a relatively large carrier mobility.[4] In combination, these two properties give rise to long diffusion lengths extending over micrometres[5] and consequently allow for efficient charge extraction in submicron thin films.

The recombination rates in LHPs are, indeed, up to 5 orders of magnitude slower than that predicted from Langevin theory for a direct semiconductor.[6] The reason behind this property is still an active topic of debate. One possibility that is gaining acceptance is that the carriers have a polaronic nature.[7] The polar character of the lattice screens the Coulomb potential of photogenerated charges, which then experience reduced scattering with other charges and lattice defects.[7,8] This effect should slow down not only recombination but also charge carrier cooling in comparison to bare charges.[9] Because of two counteracting phenomena, predicting whether the polaron formation produces an increase or a decrease in the observed carrier mobility is difficult. On the one hand, the lattice deformation associated with the polaron should produce a heavier effective mass. On the other hand, the dynamic screening of the Coulomb potential upon the formation of a large polaron will hinder the scattering of charges with lattice defects.[7] Meanwhile, hot carriers can undergo additional carrier-carrier scattering at high temperature and density, which can be screened upon polaron formation.[9] Therefore, time-resolved mobility measurements, such as those obtained through optical pump-THz probe spectroscopy, can be a useful way of detecting polaron formation and scrutinizing the hot carrier dynamics. Several spectroscopic studies of hot photocarriers have recently been reported[10–14] along with theoretical studies focusing on the slow rate of this cooling.[15,16] The time evolution of hot carriers is of particular interest in view of the possibility of realizing photovoltaic cells with conversion efficiencies exceeding the Shockley-Queisser limit. Indeed, several studies have reported long hot carrier lifetimes up to tens of picoseconds, opening the way to extracting them at selective contacts.[17]

When studying early charge carrier dynamics, a good time resolution is essential to reach meaningful conclusions with the application of mathematical models. This time resolution can be achieved in time-resolved terahertz spectroscopy (TRTS) using an ultrabroadband THz single-cycle pulse generated with a two-colour plasma technique.[18] In our case, such a pulse includes frequencies from 1 to 20 THz, allowing for a pulse duration as short as 200 fs instead of the ~1 ps pulse typically obtained by optical rectification in solid semiconductors (see the THz spectrum and pulse shape in SI, Fig. S1). The use of ultrabroadband pulses poses the difficulty of finding the right substrate with full transparency. Various approaches have been used in the past, such as employing Si wafers[19] and diamond as thin film substrates ,[20] or using single-crystals in reflectance mode.[21] The latter substrates, however, either give rise to strongly modulated absorption spectra and pose an excessive experimental and analytical complexity or are too costly. In addition, to study semiconductors in a state close to that of a working optoelectronic device, perovskite thin films should be studied in a transmission configuration that allows the bulk of the material to be probed rather than its surface. To address this problem, we developed a simple method to deposit LHP thin films on plastic sheets, such as high-density polyethylene (HDPE), fully transparent to our 1-20 THz pulses. These samples do not appear to be affected by the substrate (see the absorption spectrum in SI, Fig. S1).

**Results and analysis**

We aim to study the relationship between carrier-lattice interactions and carrier cooling from a mobility perspective using TRTS. Such a study can be achieved by exploiting the sensitivity of THz radiation transmission to the carrier mobility. In an optical pump-THz probe experiment applied to a semiconductor material, the pump generates charge carriers that absorb the incident THz probe pulse. The THz radiation absorptance is generally reported as the negative of the ratio between the change in the modulus of the THz electric field ($\Delta E$) upon transmission through the sample and the modulus of the incident field ($E$). In the case of thin films, when $\Delta E \ll E$, this ratio can be shown to be proportional to the photoconductivity ($\Delta\sigma$) generated by the photogenerated carriers, which is itself proportional to the carrier density ($N$) and the carrier mobility ($\mu$) (Eq. 1).[22,23]

$$\frac{-\Delta E}{E} \propto \Delta\sigma = \mu \cdot e \cdot N \qquad (1)$$

An evolution of the THz absorptance of the sample must therefore be associated with a change in either $N$ or $\mu$. With this in mind, we determined the early carrier dynamics of three LHP thin films of various compositions, namely, $CH_3NH_3PbI_3$, $CH_3NH_3PbBr_3$, and $CsPbBr_3$, using ultrabroadband TRTS. We exploited the fact that the photoexcited waveform $\Delta E(t)$ is not phase shifted with respect to the initial pulse. This characteristic is due to the short scattering lifetime in perovskites, yielding a moderate mobility. Therefore, this characteristic allowed us to follow the dynamics by concentrating on the point of highest electric field, giving a frequency-averaged response (see SI, $t$ = 0 in Fig. S2). Fig. 1 shows the charge carrier photoconductivity dynamics measured during the first picoseconds following photoexcitation at different pump wavelengths of a $CH_3NH_3PbI_3$ thin film deposited on HDPE. At resonant photon energies, a fast rise limited by the time resolution of the setup can be observed. Above resonant energies, a second slower rise becomes increasingly apparent with increasing excitation energy. In terms of Eq. 1, the fast initial rise can be assigned to an increase in the carrier density $N$ upon photoexcitation, while the latter rise is related to hot carrier evolution affecting the mobility $\mu$. Such behaviour of the mobility has been previously reported, e.g., for GaAs[24] or perovskites,[12,25] and is generally associated with the formation of hot charge carriers at different positions of the band structure $U_E(\mathbf{k})$ away from the bandgap threshold (Scheme 1). This behaviour is due to the relation between the mobility, effective mass and band curvature. Indeed, the mobility $\mu$ inversely depends on the effective mass $m^*$, while $m^*$ in turn inversely depends on the band curvature $\rho$.[26] In a simplified one-dimensional approximation, the carrier mobility $\mu$ is then proportional to the band curvature $\rho$ and the scattering time $\tau_s$ (Eq. 2).

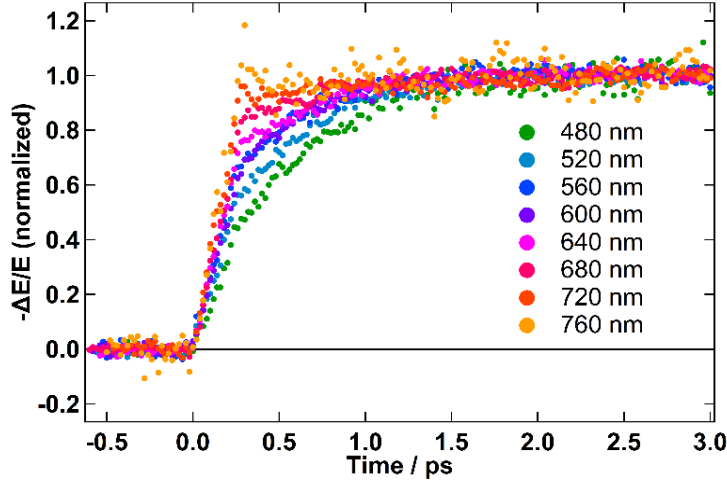

**Figure 1.** Excitation wavelength dependence of the early photoconductivity dynamics in CH$_3$NH$_3$PbI$_3$ thin films as probed by ultrabroadband TRTS. The terahertz absorptance −Δ$E$/$E$ is directly proportional to the photoconductivity Δ$\sigma$.

$$\mu = e \cdot \tau_s \cdot \frac{1}{m^*} = e \cdot \tau_s \cdot \frac{1}{\hbar^2} \cdot \frac{\partial^2 U_E}{\partial k_j^2} = \frac{e}{\hbar^2} \cdot \tau_s \cdot \rho \qquad (2)$$

Scheme 1 displays a simplified band structure of CH$_3$NH$_3$PbI$_3$, where the spin-orbit coupling has been neglected.[27] In such a situation, carriers photogenerated upon excitation in the blue region experience a significantly lower curvature of both the conduction and valence bands (positions A and A'). Hence, they must be characterized by a larger effective mass ($m^*$) and a lower mobility ($\mu$) compared to carriers photogenerated upon excitation in the red region (positions B and B').

In addition, calculations suggest that a loss of parabolicity occurs as close as one-third from the minimum point along any direction (Scheme 1).[27] This loss produces a decrease in mobility for the carriers higher up in the valley.[28,29] Thus, relatively small excess energies already produce a noticeable decrease in the initial average mobility $\mu$, even though no carriers are located in different valleys.

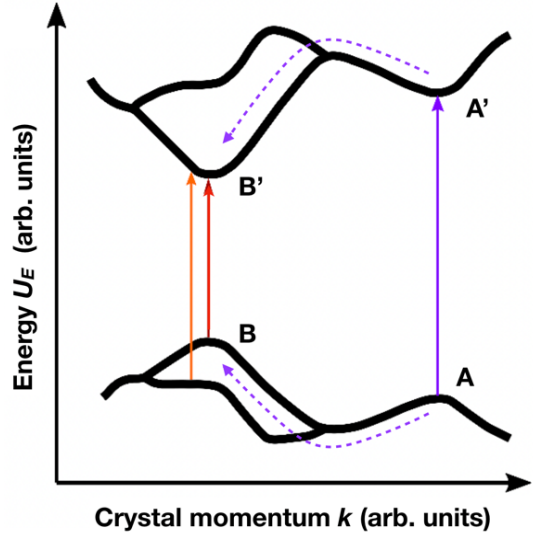

**Scheme 1.** Simplified band structure scheme based on ref. [27]. Direct transitions occurring at $\lambda_{ex}$ = 800 nm (red), $\lambda_{ex}$ = 600 nm (orange) and $\lambda_{ex}$ = 400 nm (purple) are represented by vertical arrows. The dashed arrows represent a cooling pathway of photocarriers generated upon excitation at $\lambda_{ex}$ = 400 nm.

Then, from a simple perspective, the observation of the slower rise in the THz absorptance over time can be assigned to intervalley relaxation and intravalley scattering of hot carriers with phonons towards the bottom of the valley where the effective mass accounts for the maximum mobility. In order to model such a process, we first consider a simple situation with low mobility hot carrier ($N_H$) and high mobility cold carrier ($N_C$) populations. This simplified model (see details in SI, section S3a) gives rise to a single exponential time profile of the transient photoconductivity signal (Eq. 3):

$$S(t) = 1 - B \cdot \exp\left(-\frac{t}{\tau}\right) \tag{3}$$

where $B$ is the amplitude and $\tau$ is the lifetime. In such a simplified model, carriers can only be either hot or cold. However, a distribution of carriers with different excess energies and, thus, mobilities characterized by a certain temperature exists in reality. In addition, one should consider the possibility of polaron formation. Despite its limitation, fitting this initial simple model can be useful in evaluating the lifetime of the process involved. Parameter $B$ in Eq. 3 is a measure of the average mobility of the ensemble with respect to the final situation. A convolution with a Gaussian of the full width at half maximum (FWHM) $w$ accounting for the instrument response function (IRF) can be used to fit the data (see SI, Eqs. S7, S8). The results of the data fitting shown in Fig. 1, with different measurements at different pump wavelengths ($\lambda$), are found in Table 1. Due to the difficulty of experimentally evaluating the cross-correlation between the pump and probe, the value of the FWHM *w* is also fitted. Nonetheless, we expect it to be *ca.* 200 fs, close to the width of our THz pulse.

**Table 1.** Results for the fitting of the simple exponential model (see SI, Eq. S8) at different pump wavelengths ($\lambda_{exc}$) for $CH_3NH_3PbI_3$.

| $\lambda_{exc}$ / nm | B / - | τ / ps | w / ps |
|---|---|---|---|
| 480 | 0.78 | 0.50 | 0.17 |
| 520 | 0.63 | 0.46 | 0.20 |
| 560 | 0.51 | 0.39 | 0.18 |
| 600 | 0.48 | 0.41 | 0.22 |
| 640 | 0.38 | 0.44 | 0.19 |
| 680 | 0.22 | 0.45 | 0.21 |

As expected, the amplitude *B* increases with excess energy owing to the higher initial concentration of hot carriers. Surprisingly, however, the lifetimes from the measurements are fairly constant in the 480-680 nm excitation wavelength range, with values between 400 and 500 fs. Above 680 nm, the rise is almost negligible. The lack of an increase in the lifetime is surprising if we assign it to carrier cooling, as the higher the initial carrier temperature is, the longer cooling down should generally take.[13] If the THz absorption is a direct probe of the average carrier temperature, then this would indeed be the case. However, the exponential fits with a constant lifetime indicate a different process that links high and low mobility carriers. The relaxation of hot charge carriers includes two initial steps: a) thermalization, in which the hot carriers equilibrate towards a Fermi-Dirac distribution, characterized by a certain temperature higher than that of the lattice, and b) cooling, in which the hot carriers interact with the lattice through carrier-phonon inelastic scattering with longitudinal optical (LO) phonons to shed their extra energy.[17] The former step has been reported to occur in the 10 to 100 fs interval,[30] below our actual time resolution, while the latter step can actually be studied with TRTS. LO phonons are considered to be involved not only in the relaxation of hot carriers[13] but also in polaron formation.[31] Recently, time-resolved spectroscopy techniques other than TRTS have revealed that the polaron formation time for perovskites is on the order of hundreds of fs.[31] As a type of carrier-phonon coupling, the appearance of polarons should be kinetically determined by the emergence of phonon-carrier interactions. In ref. [31], the perovskites under study were $CH_3NH_3PbBr_3$ and $CsPbBr_3$. For these materials, we obtained, on average, low mobility lifetimes of 320 and 650 fs, very close to the reported polaron formation times of 290 and 700 fs, respectively. Note that these lifetimes are subject to a certain variability depending on the sample. Interestingly, for some of the measurements reported in the literature, the observed process did not involve any cooling since a resonant excitation was used.

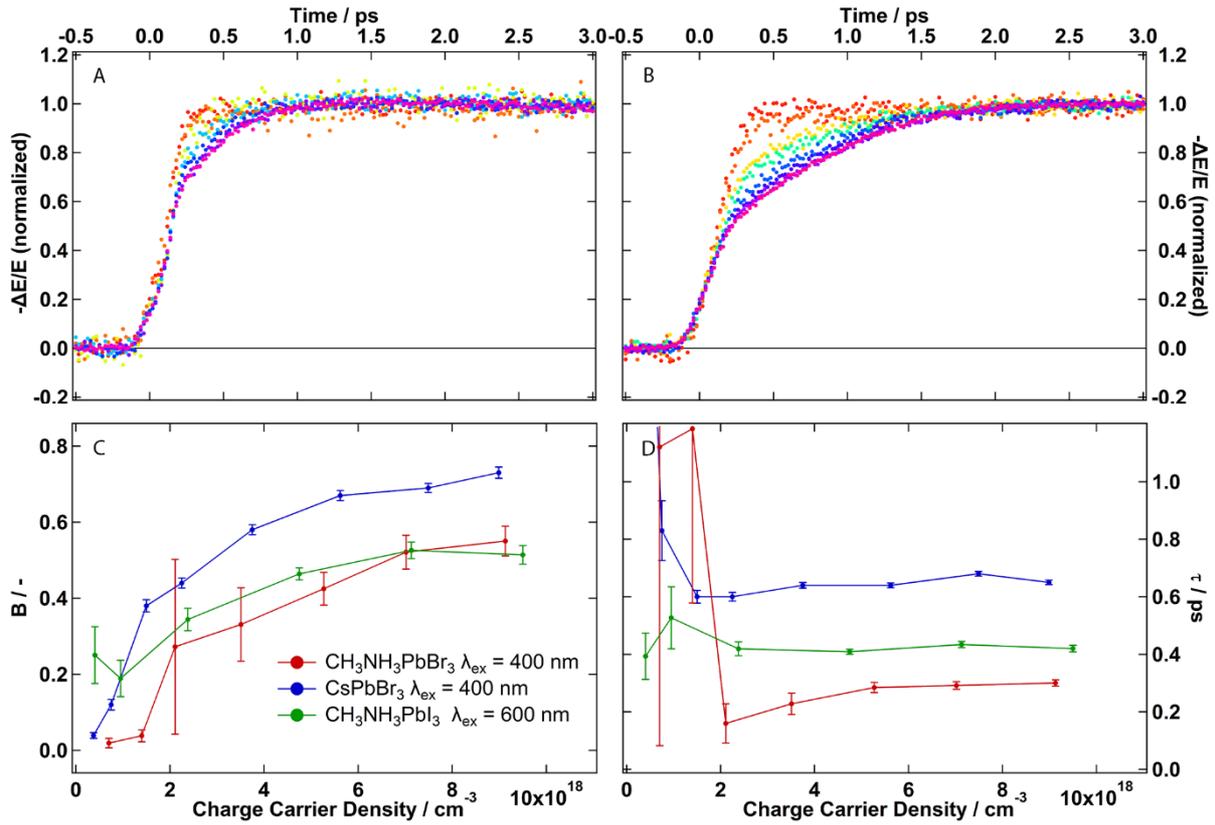

**Figure 2.** Fluence dependence of the rise in photoconductivity for A) $CH_3NH_3PbBr_3$ and B) $CsPbBr_3$. Note that both materials present a similar fluence dependence, while the lifetime is noticeably longer in the fully inorganic material. C) The amplitude *B* from the fit is shown to be the main parameter that varies with fluence, while D) the time constant $\tau$ saturates earlier. A similar behaviour is found for $CH_3NH_3PbI_3$, as seen in C) and D).

A recent publication reported a direct observation of polaron formation with TRTS.[12] However, we have to dispute the validity of the conclusions drawn in this article, due to the data treatment approach used (see discussion in SI, section S2).

From the existing experimental evidence, polaron formation in the form of a dynamic screening process has been suggested to be in competition with carrier cooling.[9] Indeed, this dynamic screening process could result in polaron formation with the observed lifetimes. In Fig. 1, we show measurements for relatively high charge carrier densities, above $10^{18}$ cm$^{-3}$. At these densities, a slow cooling process is expected to be observed due to the phonon bottleneck typically observed in these materials.[32] Thus, if two competing processes occur, we can hypothesize that, at low densities, carrier cooling to the lattice temperature is sufficiently fast to occur before polaron formation, while at higher densities, the opposite is true. To test this hypothesis, we perform fluence-dependent measurements, as shown in Fig. 2.

Indeed, the data show an increase in the magnitude of the mobility rise with increasing fluence. Our data do, nonetheless, suggest an intriguing possibility: the formation of polarons does not produce any change in mobility if the carriers are already cold. When exciting the samples at resonant energies or at low fluences, the slower rise is not observed (Figs. 1, 2 and SI, Fig. S3). This result is intriguing because polaron formation should have mobility changes associated with it due to a) an increase in the effective mass while carrying the lattice deformation and b) a decrease in scattering with defects.[7] These effects may compensate for each other for cold carriers. Moreover, the fact that a rise with the polaron formation lifetime is observed at high fluences over the bandgap indicates that the dynamic screening process does produce a substantial increase in the hot carrier mobility. The dynamic screening may possibly reduce the carrier-carrier scattering present at high carrier temperatures. Thus, the mobility would increase upon hot polaron formation, unlike with cold polaron formation. Furthermore, the hot polarons present a much slower cooling, which prevents us from observing any further dynamics due to the strong Coulomb screening of the lattice.[9,33,34] We can thus propose the model depicted in Scheme 2.

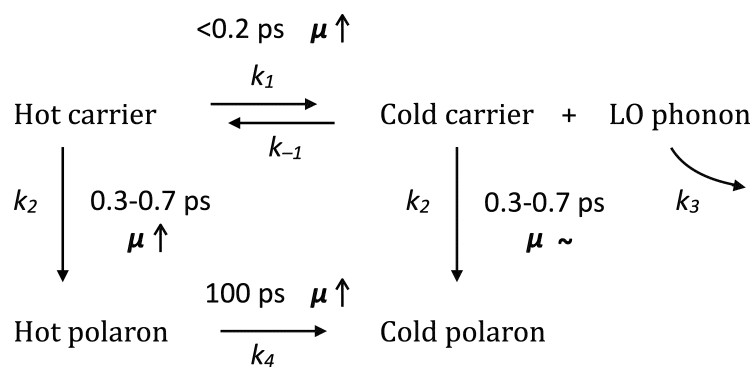

**Scheme 2.** Cooling and dynamic screening competition model. The different processes induce a change in the mobility in the direction represented by the arrows next to $\mu$. The carrier cooling and hot polaron formation processes induce a stronger increase in the mobility in the time window analysed herein.

According to this model, hot carriers turn into cold ones through phonon emission. This step is considered reversible, introducing a hot phonon bottleneck. The phonons decay at a certain rate such that a phonon population builds up at high hot carrier densities, decreasing the overall rate of carrier cooling. In addition, this cooling process competes with the polaron formation process through dynamic screening. The latter process has a certain lifetime depending on the perovskite composition.

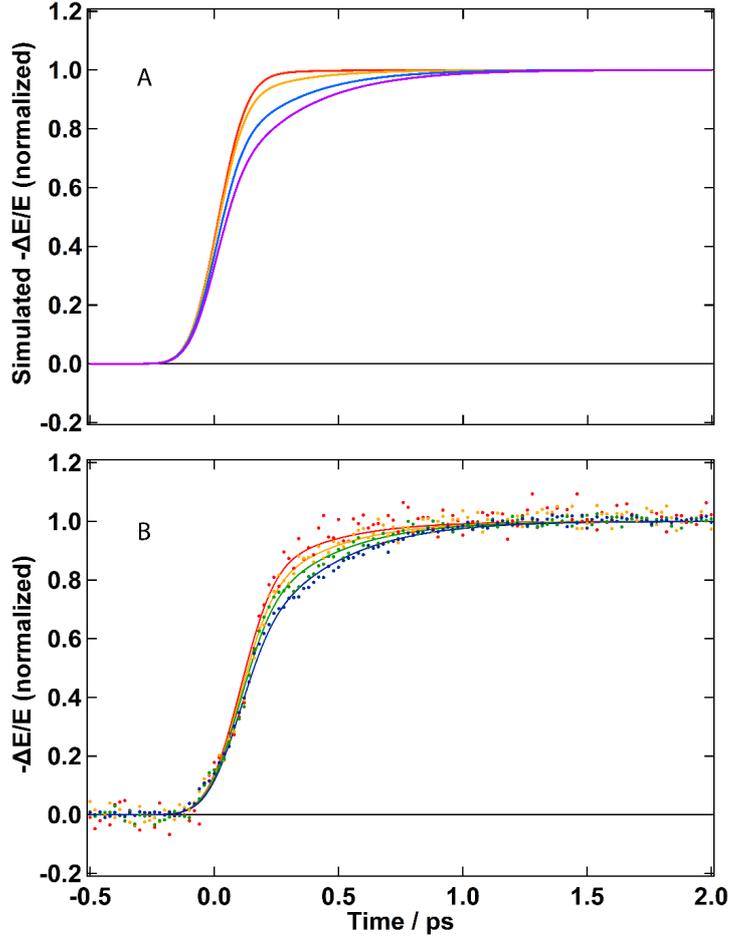

**Figure 3.** A) Simulated signals for $CH_3NH_3PbBr_3$ at different charge carrier densities following the model of Scheme 2 (see details in SI, section S3b). B) Model global fittings to $CH_3NH_3PbBr_3$ data at 18 µJ/cm² (red), 31 µJ/cm² (yellow), 46 µJ/cm² (green), and 80 µJ/cm² (blue), where all parameters are common and the initial carrier density is proportional to the fluence multiplied by a common fitted value.

As the initial hot carrier density increases, the cooling process is slowed and hot polaron formation is favoured, giving rise to very long hot (polaronic) carrier lifetimes. From our observations, we ascribe the strong increase in mobility to bare carrier cooling and hot polaron formation. Once again, we use a simplified model that does not consider carrier and phonon energy distributions with a certain temperature. Nonetheless, this model appreciably follows the excitation fluence and wavelength dependencies that we observe. We justify not using a temperature model[13] by the fact that the function relating temperature and mobility is difficult to obtain and we can hardly assume that it is linear. Moreover, using such a model would overcomplicate the computation

As shown in Fig. 3, we can numerically solve the differential equations in the model in order to simulate the signal at different carrier densities (see details in SI, section S3b). These results show the same trends as the measurements in Fig. 2A. Furthermore, a global fitting of several datasets can be carried out in which all the parameters are common except for the initial charge carrier concentration, which can be fixed to a value proportional to the relative fluence used. The good quality of the fits supports the general validity of the model, which follows the observed trend with fluence. Unfortunately, the complexity of the model prevents meaningful values from being obtained for each parameter in free fits, and many of the parameters have to be prefixed and guessed. Nonetheless, the importance lies in the general behaviour and trends with fluence and wavelength that arise from the competing cooling and dynamic screening processes. The results of the fitting procedure are shown in Table 2. We obtain a time constant for polaron formation $1/k_2$ = 0.4 ps. Large values of the kinetic constants $k_1$ and $k_{-1}$ are guessed, since the equilibrium between hot carriers, on the one hand, and cold carriers and LO phonons, on the other hand, (Scheme 2) must be reached within the time resolution of the setup. The rate constants for phonon decay ($k_3$) and polaron cooling ($k_4$) are taken from the literature.[9,13] The fitted value of the parameter $a$ = 0 (SI, Eq. S14) means that hot carriers do not contribute to the transient conductivity signal.

**Table 2.** Results of the global fitting of the complex model to experimental data as seen in Fig. 2B for $CH_3NH_3PbBr_3$. The parameter $w$ stands for the FWHM of the convoluted Gaussian, a measure of the time resolution of the setup. The initial hot carrier density $N_{hc}(0)$ is given in arbitrary density units (adu). The same units are used in the dimensions of the second order rate constant $k_{-1}$.

| Parameter | *Fitted value* |
|---|---|
| $a$ | 0 |
| $N_{hc}(0)$ | 0.3 adu |
| $w$ | 0.24 ps |
| $k_1$ | 200 ps$^{-1}$ (guessed) |
| $k_{-1}$ | 100 adu$^{-1}$ ps$^{-1}$ (guessed) |
| $k_{-1} \cdot N_{hc}(0)$ | 30 ps$^{-1}$ |
| $k_2$ | 2.5 ps$^{-1}$ |
| $k_3$ | 1.67 ps$^{-1}$ (literature) |
| $k_4$ | 0.01 ps$^{-1}$ (literature) |

Further information on these materials can be extracted from a comparison between different compositions in terms of cations and anions. One key parameter is affected: the dynamic screening lifetime, on which both anions and cations have an effect. $CsPbBr_3$ is the material presenting the longest lifetime, suggesting that the extra degrees of freedom from the organic cations play a role in speeding up the process. Last, low temperature measurements show an increase in the lifetime of the mobility rise accompanied by an increase in the total terahertz absorption (see SI, Fig. S4). The latter phenomenon is known in perovskites, where the polaron mobility is limited by polaron-phonon scattering.[19] Thus, at lower temperatures, the mobility is higher due to the lower population of phonons. The former phenomenon indicates a role of the phonon density in the polaron formation process.

**Conclusions**

Hot carrier dynamics LHPs of different compositions have been analysed from a THz mobility perspective. The results are compatible with competing cooling and dynamic screening processes. A model was proposed and applied to explain the observed fluence dependence and constant lifetime of the mobility rise. The composition was found to mainly affect the time constant of the dynamic screening process. In this regard, the $Cs^+$ inorganic cation was found to produce longer dynamic screening than the organic counterparts for a given halide composition. This result is probably due to a decrease in the degrees of freedom that allow for polarization of the lattice in the presence of a charge carrier. Last, low temperature measurements hint at implications of the phonon densities in the dynamic screening process.

**Methods**

Sample preparation

Thin film samples were prepared by spin-coating using an antisolvent method[35] on fully THz transparent HDPE substrates (1 × 15 × 15 mm). The substrates were cleaned using a sonicating bath in a Hellmanex® solution and subsequently rinsed with ultrapure water, acetone, ethanol and high purity methanol. Directly afterward, the substrates were submitted to an oxidative plasma treatment for 90 min. This treatment allowed a better wettability of the apolar polymer with polar solutions through the creation of oxidized polar groups on the surface. Precursor solutions (1 mmol in 1 mL of DMSO) were prepared for deposition. In the case of $CsPbIBr_3$, due to the low solubility, the saturated solution of the mentioned mixture was used. Spin-coating was carried out as follows: a sufficient amount of solution was extended over the substrate. The sample was spun at 1000 rpm for 10 s and subsequently at 6000 rpm for 30 s. Ten seconds before the end, 200 μL of chlorobenzene was poured onto the spinning substrate to act as an antisolvent and improve crystallization. Last, the sample was annealed at 100 °C to obtain a perovskite layer of the characteristic colour (see spectral characterizations of the films in SI, Fig. S1).

## TRTS spectroscopy

TRTS measurements were carried out on a previously described laser spectrometer,[36] with a few modifications. Three beams were split from the fundamental output (45 fs pulse duration, $\lambda$ = 800 nm, 1 kHz repetition rate) of an amplified Ti:sapphire laser (Libra, Coherent) and used for the TRTS experiments. The first beam was employed to pump a white light-seeded optical parametric amplifier (OPerA Solo, Coherent) that provided the pump pulses for the pump–probe experiment at tuneable wavelengths. Alternatively, the same beam could be diverted to obtain 400 nm pulses through second harmonic generation in a BBO crystal. The more powerful of the two remaining beams (390 µJ per pulse) was used to generate the probe beam, consisting of a train of short and broadband THz pulses, through a two-colour plasma method:[37] The beam was focused with a fused silica lens ($f$ = 75 mm), and the second harmonic was generated with a 100 µm BBO crystal. At the focal point, the electric field of the two-colour beam was sufficiently strong to form a plasma filament in nitrogen that radiated a broadband THz pulse (200 fs, 1–20 THz) that was subsequently collimated and focused with parabolic gold mirrors onto the sample. To maximize THz generation, a dual wavelength waveplate was used immediately after the BBO crystal to obtain fundamental and second harmonic beams of equal polarization after type 1 phase matching.[38] The transmitted beam went through two additional parabolic mirrors towards a homemade ABCD (air-biased coherent detection) detector.[39] Silicon wafers were used to filter the visible light from the THz generation and pumping. The remaining beam (40 µJ/pulse) was used as a gate for the detection, generating a second harmonic signal proportional to the THz electric field measured with a PMT (PMM01, Thorlabs). The SHG process was carried out in an enclosed box with TPX® windows, where the atmosphere was replaced with butane gas.[40] This setup allowed an increase in the sensitivity at the expense of frequencies in the 15-20 THz range and marginally increased the time resolution.

## Calculations

The data were fitted to the model using Wolfram *Mathematica* software. The differential equations were written following the model depicted in Scheme 3 and numerically solved for an interval covering the data. The signal was simulated as a normalized and weighted sum of the different carrier densities with the condition that it equals zero for times earlier than $t_0$. The result was numerically convoluted with a Gaussian function to account for the IRF. An interpolation function was fitted to the results to calculate the value at any times inside the calculated interval. A small time step of 0.005 fs was used for the calculation, which was shown to be sufficient to obtain results identical to those with simpler functions for which an analytical solution is possible. The function was programmed such that once the interpolation function had been obtained, for a given set of parameters, it could be called to obtain the values at different times. Therefore, the resulting overall function could be used not only to simulate the signal but also to fit the data.

## Data availability

The data that support the plots within this paper and other findings of this study are available from the corresponding author upon reasonable request.

**Acknowledgements**

Financial support by the Swiss National Science Foundation (SNSF, Grant No. 200021_175729) and the National Center of Competence in Research "Molecular Ultrafast Science and Technology" (NCCR-MUST), a research instrument of the SNSF, is gratefully acknowledged.

## Authors contributions

ABC and JEM conceived the project. JEM supervised the project and designed the experiments with ABC. ABC prepared the samples and carried out all optical and THz measurements with the assistance of ARW and JMMN. ABC conceived the kinetic competition model, ran the simulation and fitted the data. AA, AAP and ABC built and tested the time-resolved ultrabroadband THz spectrometer and carried out preliminary measurements. ABC wrote the manuscript with JEM. All other authors discussed the results and edited the manuscript.

## Competing interests

The authors declare no competing interests.

## Supplementary Information

Details on the selection of a suitable polymer substrate, the sample preparation and the optical characterization – Ultrabroadband THz pulse and spectrum characterization – Discussion on the minimum time resolution necessary to evidence dynamic screening and the particular case of ref. [12] – Details of the mathematical models – Supplementary experimental data: a) Fluence dependence of the THz absorptance signal at resonant excitation for $CH_3NH_3PbBr_3$ and $CsPbBr_3$, and b) Low temperature measurements of the early dynamics of the THz absorptance in $CH_3NH_3PbBr_3$.

# Hot Carrier Mobility Dynamics Unravel Competing Sub-ps Cooling Processes in Lead Halide Perovskites

Andrés Burgos-Caminal, Juan Manuel Moreno-Naranjo, Aurélien René Willauer, Arun Abi Paraecattil, Ahmad Ajdarzadeh, and Jacques-E. Moser *

Photochemical Dynamics Group, Institute of Chemical Sciences and Engineering, École Polytechnique Fédérale de Lausanne, 1015 Lausanne, Switzerland

* Corresponding author. Email: je.moser@epfl.ch

# Supplementary Information

### S1. Polymer substrate

When choosing a polymer to be used as a substrate in time-resolved terahertz spectroscopy (TRTS), minimization of the vibrational transitions that can produce strong absorption bands in the THz region is important. The first polymers to discard are those containing polar groups, such as poly(methyl methacrylate) (PMMA) or poly(ethylene terephthalate) (PET), or even any heteroatom, such as polytetrafluoroethylene (PTFE, Teflon™),[1] since they involve complete or partial absorption of the 1-20 THz spectrum, especially when using millimetre thick substrates. Therefore, we considered simple polymers, such as polypropylene (PP), polymethylpentene (PMP, TPX™), and high-density polyethylene (HDPE). In addition, branching (of a few carbon atoms) will generally lead to higher degrees of freedom and thus more vibrational modes to be excited. Indeed, HDPE, the simplest polymer, was found to be the most transparent up to a frequency of 20 THz and was chosen as the substrate to support lead halide perovskite (LHP) films in our experiments. TPX™ is an interesting material that can be used in windows for special applications in TRTS due to its transparency in both the THz region, up to 12 THz, and the visible region.

Thin films of LHPs composed of various anions and cations were obtained on 1 mm HDPE sheets through a spin-coating procedure, identical to that generally used on glass or similar supports. When spin-coating solutions to form LHP thin films, one of the key parameters, and the one that mainly varies between HDPE and glass, is the wettability of the surface to the polar solvent (DMSO). HDPE is an apolar material that repel polar solvents, making the spin-coating technique impracticable. Indeed, the spin-coating technique is based on the formation of a thin film of solution when a spinning force is applied. However, if poor interaction occurs between the solvent and the substrate, the solution just slips off. To solve this issue, a surface treatment must be applied to the HDPE surface to endow it with a polar nature. This nature was achieved through an oxidative plasma treatment. Such an approach generates oxidation products on the surface of the polymer that constitute polar groups, highly increasing the wettability to polar solvents.[2] The resulting LHP films are polycrystalline and have photophysical properties similar to the films prepared on other substrates (Fig. S1).

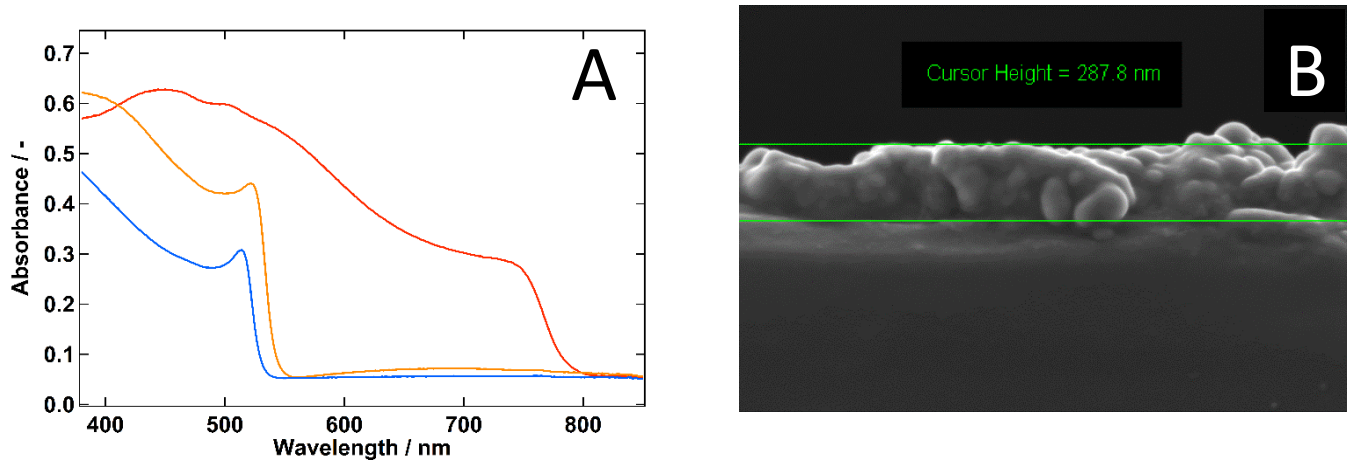

**Figure S1.** A) Absorbance spectrum of thin films of $CH_3NH_3PbI_3$ (red), $CH_3NH_3PbBr_3$ (orange) and $CsPbBr_3$ (blue) spin-coated on a high-density polyethylene (HDPE) substrate. B) SEM cross section picture of the $CH_3NH_3PbI_3$ film sample. The thickness of the perovskite layers is between 250 and 300 nm.

**S2. Ultrabroadband THz pulses, which have a decisive advantage over previous TRTS measurements of the hot carrier dynamics in LHPs**

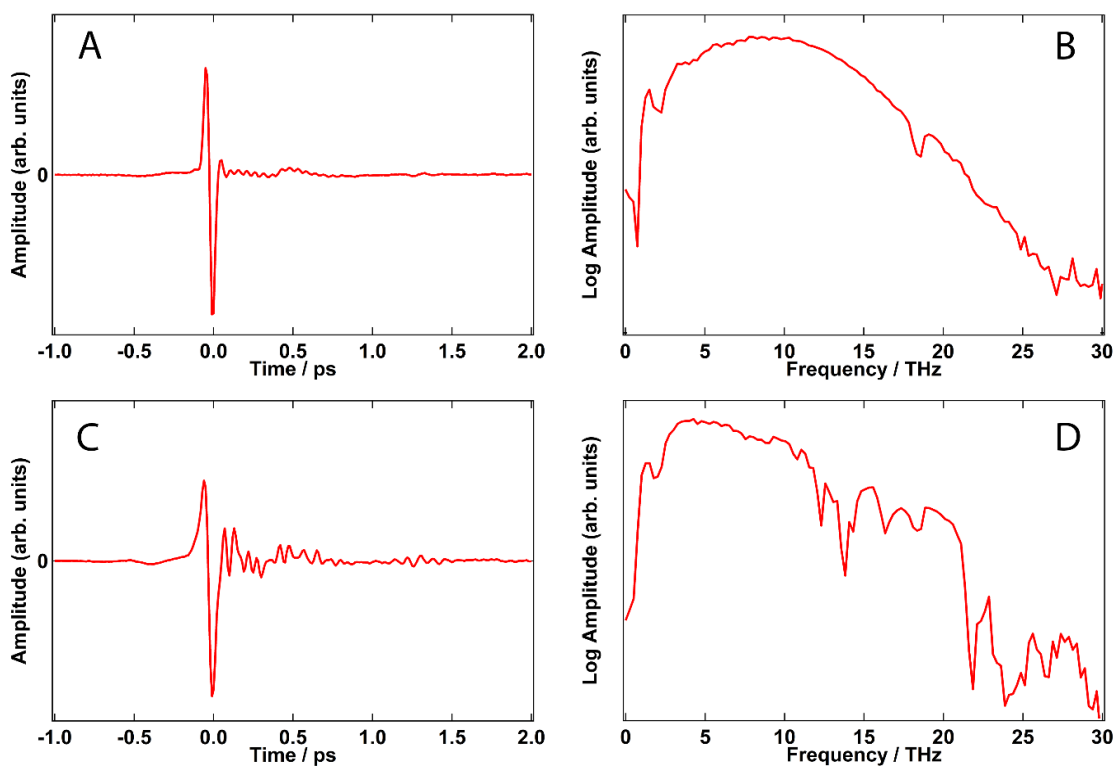

**Figure S2.** Typical pulse profiles obtained by ultrabroadband THz generation and analysis by plasma photonics as used in the present study. A) Temporal profile of the pulse detected in $N_2$. B) FT frequency profile obtained from the pulse shown in A. C) Temporal profile of the pulse detected in butane with TPX windows and D) its corresponding FT frequency profile. The observed dip at 18.5 THz corresponds to the phonon absorption band of the Si wafers used as filters.[3,4]

One of the main advantages of using an ultrabroadband TRTS spectrometer is the improved time resolution owing to the generation of shorter THz pulses. Measurements similar to those presented in this manuscript have previously been reported by Bretschneider *et al.*[5] However, regrettably, those measurements were carried out with a classical spectrometer based on optical rectification in ZnTe crystals, where the obtained THz pulses have a duration ≥ 1 ps. This approach produces a considerable loss in time resolution, which can be easily observed when comparing the results with our own measurements. A clear difference can be observed by comparing the trace corresponding to an excitation wavelength of 760 nm, with no excess energy, presented in Fig. 1 of the present manuscript with that appearing in Fig. 2A of ref. [5]. While the former trace shows a fast rise with a time constant of ca. 200 fs, the latter trace displays a rise that takes up to 1 ps to complete. Both measurements were applied to $CH_3NH_3PbI_3$ films in similar conditions, and the difference cannot be explained in terms other than a difference in the time resolution due to the cross-correlation between the pump and probe pulses. As a result, the model fitted in the cited reference, with the parameters obtained, does not show any agreement with our data because it considers a slow exponential rise to be present, even when no excess energy is employed, which is a rise that we do not observe. Therefore, the reported rise must come from the difference between the widths of the instrument response function (IRF) employed in their model and the actual IRF of their data. Thus, this fact makes us doubt the validity of the conclusions obtained from their data analysis.

**S3. Mathematical models**

a) Simple model

As a first attempt, we can consider a simple model taking into account only hot and cold carriers, whose populations $N_H$ and $N_C$, respectively, evolve according to :

$$\frac{dN_H}{dt} = -\frac{1}{\tau} \cdot N_H(t) \tag{S1}$$

$$\frac{dN_C}{dt} = \frac{1}{\tau} \cdot N_C(t) \tag{S2}$$

With the initial conditions are $N_H(0) = N^0{}_H(0)$ and $N_c(0) = 1 - N^0{}_H$, we obtain:

$$N_H(t) = N_H^0 \cdot \exp\left(\frac{-t}{\tau}\right) \tag{S3}$$

$$N_C(t) = 1 - N_H^0 \cdot \exp\left(\frac{-t}{\tau}\right) \tag{S4}$$

The assumption that both hot and cold carriers will give a signal, albeit proportionally different to the concentration due to the different mobilities, is reasonable. Thus, for a normalized dataset, we can define the following model for the sample conductivity $S(t)$ :

$$S(t) = 1 - N_H^0 \cdot \exp\left(\frac{-t}{\tau}\right) + b \cdot N_H^0 \cdot \exp\left(\frac{-t}{\tau}\right) \tag{S5}$$

where $b$ is a proportionality constant due to the different mobilities. However, $b$ and $N^0_H$ are mutually dependent and cannot be determined through a fit. Thus, further simplification of the model with the substitution $B = (1-b)N^0_H$ is necessary, yielding a single exponential :

$$S(t) = 1 - B \cdot \exp\left(-\frac{t}{\tau}\right) \tag{S6}$$

If $b$ or $N^0_H$ can be determined from a separate method, then the other parameter can be calculated.

After convoluting the exponential equation (Eq. S6) with a Gaussian function of the form:

$$\frac{2}{w}\sqrt{\frac{\ln(2)}{\pi}} \cdot \exp\left(-\frac{4\ln(2) \cdot t^2}{w^2}\right) \tag{S7}$$

representing the cross-correlation of the pump laser and the THz probe pulses, and where $w$ is the full width at half maximum (FWHM), we obtain the following equation to be fitted to the observed dynamics:

$$S(t) = \frac{1}{2}\left(1 + \text{erf}\left[2(t-t_0)\sqrt{\frac{1}{w^2}} \cdot \sqrt{\ln(2)}\right] - B \cdot \exp\left[\frac{w^2\left(1 - \frac{16(t-t_0) \cdot \tau \cdot \ln(2)}{w^2}\right)}{16\,\tau^2 \cdot \ln(2)}\right] \cdot \text{erfc}\left[\frac{1 - \frac{8(t-t_0) \cdot \tau \cdot \ln(2)}{w^2}}{4\text{tc}\sqrt{\frac{1}{w^2}} \cdot \sqrt{\ln(2)}}\right]\right) \tag{S8}$$

b) Competition model

The competition model is obtained by numerically solving the below system of five differential equations (Eqs S9-S13), where $N_{hc}$, $N_{cc}$, $N_{loph}$, $N_{hp}$, and $N_{cp}$ correspond to the populations of the hot carriers, cold carriers, LO phonons, hot polarons, and cold polarons, respectively. The rate constants $k_1$, $k_{-1}$, $k_2$, $k_3$, and $k_4$ are characterise the processes depicted in Scheme 2.

$$\frac{dN_{hc}(t)}{dt} = -k_1 \cdot N_{hc}(t) + k_{-1} \cdot N_{cc}(t) \cdot N_{loph}(t) - k_2 \cdot N_{hc}(t) \tag{S9}$$

$$\frac{dN_{cc}(t)}{dt} = k_1 \cdot N_{hc}(t) - k_{-1} \cdot N_{cc}(t) \cdot N_{loph}(t) - k_2 \cdot N_{cc}(t) \tag{S10}$$

$$\frac{dN_{loph}(t)}{dt} = k_1 \cdot N_{hc}(t) - k_{-1} \cdot N_{cc}(t) \cdot N_{loph}(t) - k_3 \cdot N_{loph}(t) \tag{S11}$$

$$\frac{dN_{hp}(t)}{dt} = k_2 \cdot N_{hc}(t) - k_4 \cdot N_{hp}(t) \tag{S12}$$

$$\frac{dN_{cp}(t)}{dt} = k_2 \cdot N_{cc}(t) + k_4 \cdot N_{hp}(t) \tag{S13}$$

The conductivity signal *S(t)* is finally approximated by considering that hot polarons, cold carriers, and cold polarons have the same mobility:

$$S(t) = N_{hp}(t) + a \cdot N_{hc}(t) + N_{cp}(t) + N_{cc}(t) \tag{S14}$$

where $N_{hp}(0) = 0$ and $N_{cp}(0) = 0$.

## S4. Supplementary data

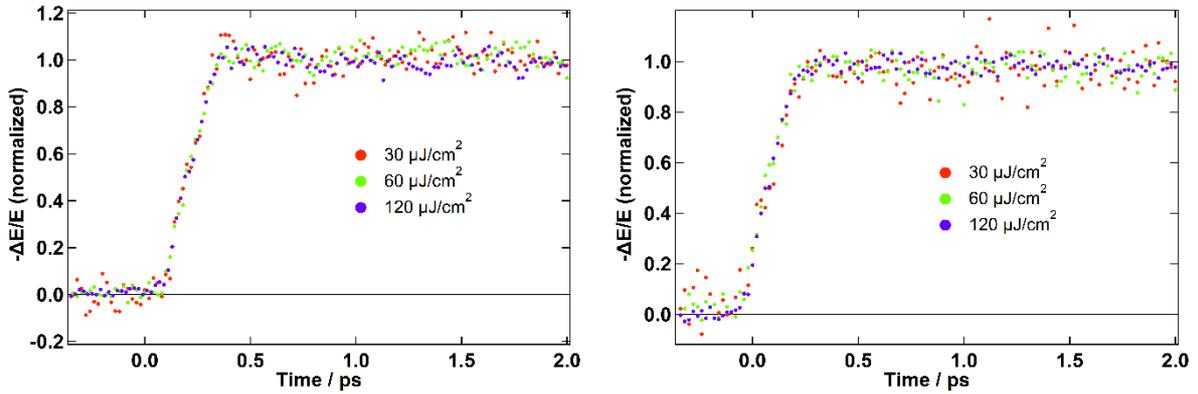

**Figure S3.** Fluence dependence of the THz absorptance signal at resonant excitation for CH$_3$NH$_3$PbBr$_3$ ($\lambda_{exc}$ = 520 nm, left) and CsPbBr$_3$ ($\lambda_{exc}$ = 510 nm, right). No dependence of the time evolution is observed.

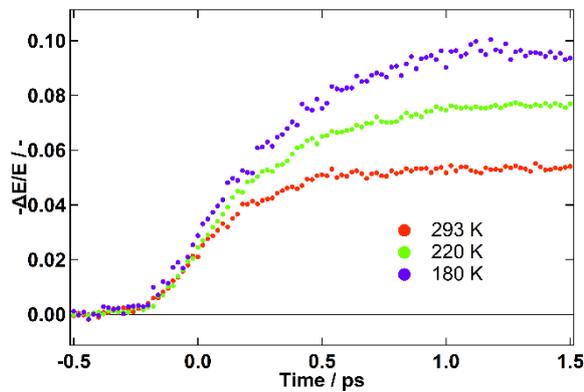

**Figure S4**. Low temperature measurements of the early dynamics of the THz absorptance in CH$_3$NH$_3$PbBr$_3$ ($\lambda_{exc}$ = 400 nm).

**References of the SI**